\documentclass[twocolumn]{aastex6}
\usepackage{graphicx}
\usepackage{amssymb}
\usepackage{amsmath}
\usepackage{hyperref}
\usepackage{ulem}

\def\cxo{{\sl CXO\ }}

\newcommand{\ndash}{--}

\def\gtrsim{\mathrel{\hbox{\rlap{\hbox{\lower4pt\hbox{$\sim$}}}\hbox{$>$}}}}
\def\lesssim{\mathrel{\hbox{\rlap{\hbox{\lower4pt\hbox{$\sim$}}}\hbox{$<$}}}}
\def\gtrsim{\mathrel{\hbox{\rlap{\hbox{\lower4pt\hbox{$\sim$}}}\hbox{$>$}}}}
\def\farcs{\hbox{$.\!\!^{\prime\prime}$}}

\def\fdegm{\hbox{$.\!\!^{\circ}$}}

\begin{document}

\title{Evolution of the extended X-ray emission from the PSR\,B1259--63/LS\,2883 binary in the 2014\ndash2017 binary cycle }
\author{Jeremy Hare\altaffilmark{1,2,3}, Oleg Kargaltsev\altaffilmark{1,2}, George Pavlov\altaffilmark{4}, Paz Beniamini\altaffilmark{1,2}}
\altaffiltext{1}{Department of Physics, The George Washington University, 725 21st St. NW, Washington, DC 20052}
\altaffiltext{2}{The George Washington Astronomy, Physics, and Statistics Institute of Sciences (APSIS)}
\altaffiltext{3}{Space Sciences Laboratory, 7 Gauss Way, University of California, Berkeley, CA 94720-7450, USA}
\altaffiltext{4}{Department of Astronomy $\&$ Astrophysics, Pennsylvania State University, 525 Davey Lab, University Park, PA 16802, USA}
\email{jhare@berkeley.edu}

\begin{abstract}
 We have performed a series of {\sl Chandra X-ray Observatory} observations of the gamma-ray binary LS 2883, which is comprised of a young pulsar (PSR B1259--63) orbiting a massive Be star with a period of 1236.7 days. The system was observed in 5 epochs, spanning a range from 352 to 1175 days after the periastron passage on 2014 May 4. The observations confirmed the recurrent nature of the high-speed ejecta that appear as an extended X-ray structure (clump)
 moving away from the binary. Compared to the results of the previous monitoring campaign (between the 2010 and 2014 periastron passages), this time we find evidence suggesting that the clump is accelerated to a projected velocity $ v_{\perp}\approx0.15c$ with an acceleration $a_{\perp}=47\pm2$ cm s$^{-2}$ (for uniformly accelerated motion), assuming that it was launched near periastron passage.  The observed X-ray properties of the clump are consistent with synchrotron emission from pulsar wind particles accelerated at the interface between the pulsar wind and the clump. We have also performed contemporaneous observations with the {\sl Hubble Space Telescope}, which are used to set an upper limit on the optical flux of the extended emission.
\end{abstract}
\keywords{pulsars: individual (B1259-63) – X-rays: binaries – X-rays: individual (LS 2883)}

\section{Introduction}
\label{intro}
High-mass $\gamma$-ray binaries (HMGBs) consist of a massive early-B or late-O type star and either a neutron star (NS) or a black hole (BH). Only 7 such sources have been discovered to date. In two of these systems, 
LS\,2883/PSR\,B1259--63 (B1259 hereafter) and MT91\,213/PSR\,J2032+4127, radio pulsars were detected, while the nature of compact object is still unknown in the other systems. The most studied of these binaries is B1259. The pulsar has a spin period $P=47.8$ ms, spin-down power $\dot{E}=8.3\times10^{35}$ erg s$^{-1}$, dipolar  magnetic field $B_{\rm PSR}=3.3\times10^{11}$ G, and characteristic age $\tau_{\rm PSR}=P/(2\dot{P})=330$ kyr \citep{1992ApJ...387L..37J}. The high-mass star in this system is a fast-rotating Be star with a luminosity $L_*=6.3\times10^{4}L_{\odot}$ and mass $M_{*}=(15-31)M_{\odot}$ \citep{2011ApJ...732L..11N,2018MNRAS.479.4849M}. The companion star is believed to have an equatorial decretion disk, inclined by $\approx35^{\circ}$ to the orbital plane \citep{1995MNRAS.275..381M,2014MNRAS.437.3255S}.

B1259 has an orbital period  $P_{\rm orb}=1236.7$ days and large eccentricity $e=0.87$ \citep{1992ApJ...387L..37J}. Recently, precise astrometric radio measurements of the system have allowed \cite{2018MNRAS.479.4849M} to accurately determine the orbital parameters and annual parallax of B1259. The updated parameters for the system are an inclination $i=153^{\circ}\pm3^{\circ}$ (implying that the pulsar orbits its companion in the clockwise direction), argument of periastron $\omega=138\fdg7$, and de-projected semi-major axis $a\approx6$ AU \citep{2018MNRAS.479.4849M}. The measured  parallax places the system at a distance $d=2.6^{+0.4}_{-0.3}$ kpc, after correcting for the Lutz-Kelker bias \citep{1973PASP...85..573L,2018MNRAS.479.4849M}.

This system has been observed many times in X-rays over several binary cycles and has shown variations in the flux, photon index, and hydrogen absorbing column density, which are dependent on the orbital phase of the observation \citep{2006MNRAS.367.1201C, 2009MNRAS.397.2123C,2015MNRAS.454.1358C}. The detected X-ray emission is non-pulsed and is thought to be due to either synchrotron or inverse Compton (IC) radiation from relativistic leptons from a pulsar wind (PW) being accelerated at the shock between the stellar wind and PW \citep{1997ApJ...477..439T}.

At GeV energies B1259 has been detected as a transient source appearing after the 2010, 2014, and 2017 periastron passages, perhaps near the time of the second disk crossing \citep{2011ApJ...736L..11A,2015ApJ...811...68C,2018ApJ...863...27J,2018ApJ...862..165T}.  The GeV flares exhibit similar spectra but different variability timescales  \citep{2015ApJ...811...68C,2018ApJ...863...27J} and   are not  accompanied by simultaneous TeV or X-ray flares  \citep{2013A&A...551A..94H,2015MNRAS.454.1358C}. For instance, in 2010 and 2017  GeV emission was detected $\sim$10 days after periastron passage and then was not detected again until $\sim$40 days later, while in 2014  GeV emission was not detected until 31 days after periastron passage. These flares lasted $\sim$30-40 days before decaying back below the detection limit. The luminosity of the GeV flare seen after the 2017 periastron passages has exceeded the spin-down luminosity of the pulsar. Additionally, rapid variability (on timescales of $1.5$ minutes)  was detected for the first time during the 2017 GeV flare. 
 
 The physical origin of the GeV flares remains unknown, with various scenarios involving, e.g., IC scattering of soft photons (of different origins) off PW
electrons  \citep{2012ApJ...752L..17K,2013A&A...557A.127D, 2017ApJ...836..241T, 2018MNRAS.476..766Y}, reconnection in the pulsar wind \citep{2013ApJ...776...40M}, interactions between the PW and  clumpy stellar wind \citep{2017A&A...598A..13D}, or  synchrotron radiation  Doppler-boosted in the bow shock tail \citep{2011ApJ...736L..10T, 2012ApJ...753..127K}. Recently, a marginal ($\sim3\sigma$) detection of B1259 in its ``quiescent'' state (i.e., far from periastron) has been reported in the {\sl Fermi}-LAT $5-300$ GeV band, apparently dependent on the orbital phase \citep{2016ApJ...828...61X}.

At TeV energies,  emission from B1259 has been detected  by H.E.S.S. \citep{2005A&A...442....1A,2009A&A...507..389A,2013A&A...551A..94H} around the time of the periastron passages. Combined data from 3 subsequent binary cycles indicate  a double-peak profile of the TeV light curve reminiscent of that  seen  at  X-ray  energies \citep{2017ICRC...35..675R}. The TeV emission could be due to either IC  scattering of stellar photons off ultra-relativistic electrons \citep{2007MNRAS.380..320K,2009A&A...507..389A} or hadronic interactions between the PW  and the companion's dense decretion disk \citep{2009A&A...507..389A}.

 B1259's pulsed radio emission disappears for nearly one month near periastron, suggesting that it becomes eclipsed by the dense stellar wind and the decretion disk of the Be-type companion star \citep{2005MNRAS.358.1069J}. Non-pulsed radio emission peaks  $\approx$ 5 days before and $\approx$20 days after periastron passage and varies in relative intensity between orbital cycles.  This emission has been interpreted as synchrotron radiation from electrons accelerated by the passage of the pulsar through the decretion disk \citep{1999ApJ...514L...39B}.  The X-ray and TeV light curves also show peaks when the pulsar enters and exits the disk, $\approx 20-4$ days before periastron and $\approx 10-50$ days after periastron, respectively \citep{2015MNRAS.454.1358C,2013A&A...551A..94H}.

 Extended X-ray emission near B1259 was first reported by \cite{2011ApJ...730....2P}, who found a faint, asymmetric extension observable up to $\approx4''$ south-southwest from the location of the binary in a 26 ks {\sl Chandra X-ray Observatory} ({\sl CXO}) observation taken near apastron. During the next orbital cycle, two deeper (56 ks exposures) {\sl CXO} observations were taken 370 and 886 days after periastron, respectively. In these observations, a variable and extended ($\sim4''$) feature, nicknamed the ``clump'', was discovered and had shifted by 1\farcs8$\pm$0\farcs5 between the observations (\citealt{2014ApJ...784..124K}, K+14 hereafter). This shift allowed K+14 to measure the corresponding projected velocity $v_\perp=(0.052\pm0.015)c$ of the clump assuming $d=2.6$ kpc. To better constrain the motion of the clump, a Director Discretionary Time (DDT) {\sl CXO} observation was undertaken. This allowed \citealt{2015ApJ...806..192P} (P+15 hereafter) to further track the motion of the extended feature and calculate a projected velocity of  $v_\perp\approx(0.08\pm0.01)c$ 
 at $d=2.6$ kpc. Surprisingly, the extended feature showed no signs of decelerating and actually showed marginal signs of acceleration (albeit at a low significance). If a constant velocity is assumed, then the clump's launch time was consistent with both periastron passage and the GeV flare.

The flux of the clump faded over time, but the photon index did not soften, implying no cooling of the clump's matter (P+15). The large projected velocity, lack of spectral softening, and launch time being consistent within $\sim\pm50$ days of periaston led P+15 to suggest that the clump is launched due to an interaction of the pulsar with the decretion disk of the  Be-star. They then suggested that the clump is moving in (and possibly accelerated by) the unshocked PW, and that the X-ray emission can be understood as synchrotron radiation from the PW shocked by the collision with the clump.

Here we report on a {\sl CXO} campaign to determine if this phenomenon is repeating, by searching for a newly launched clump after the 2014 periastron passage on May 4, 2014. We also obtained {\sl CXO} and {\sl Hubble Space Telescope} ({\sl HST}) DDT observations to help constrain the clump's motion and its multi-wavelength spectral energy distribution, and to test different energy emission mechanisms. In Section \ref{obsanddr} we describe the observations and data reduction, in Section \ref{DA} we discuss the X-ray data analysis, and in Section \ref{discuss} we discuss the implications that the new observations  impose on the previously suggested models. Finally, we  summarize our findings in Section \ref{summ}.

\begin{figure}
\centering
\includegraphics[trim={100 0 0 0},width=7.0cm]{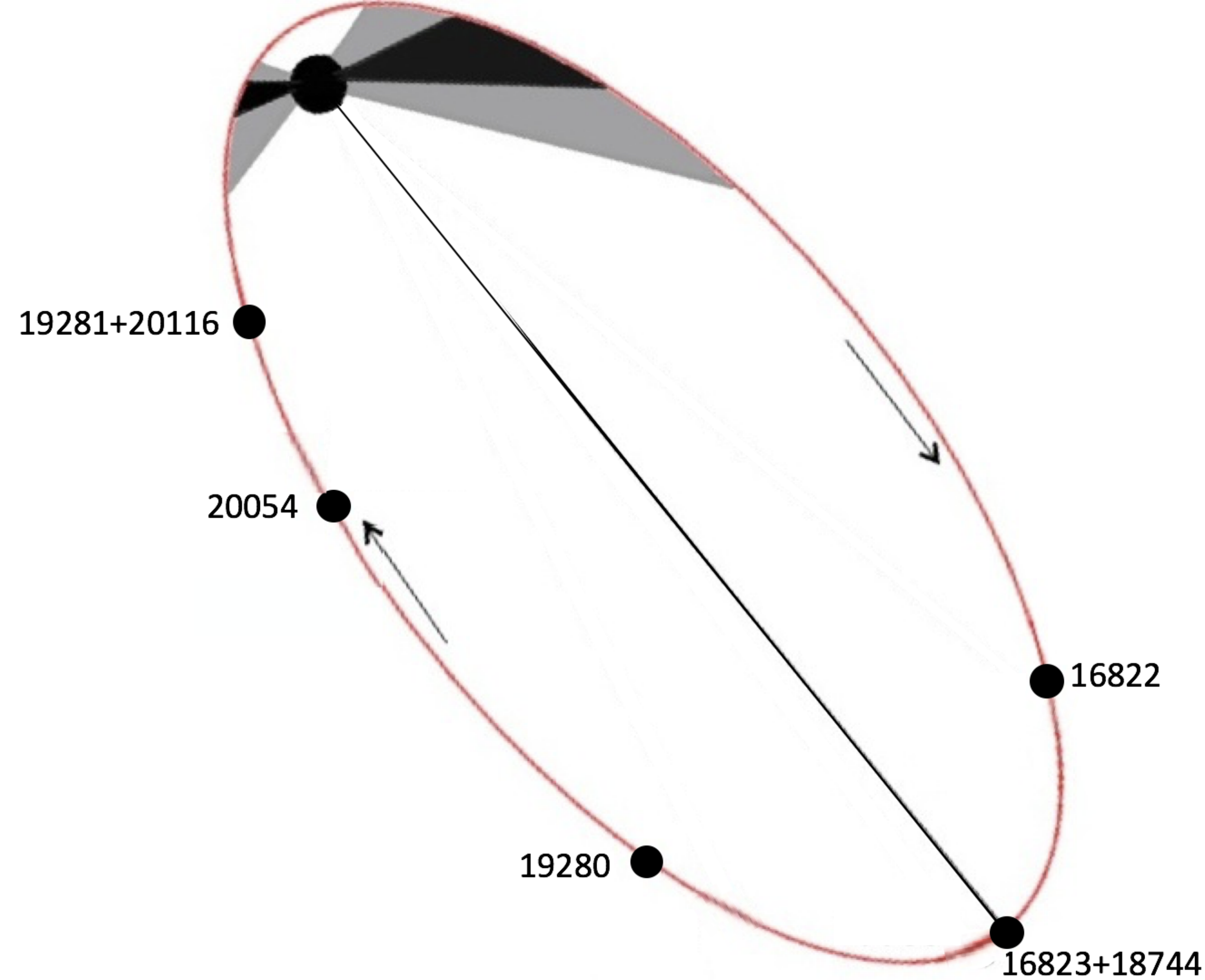}
\caption{
The orbital positions of the pulsar, marked by their ObsIDs,
at the times of 
our {\sl Chandra} ACIS observations between the 2014 and 2017 periastrons. The gray/black shaded regions show where in the orbit 
 the pulsar passes through the massive star's decretion  disk.
\label{pic1}
}
\end{figure}

\section{Observations and Data Reduction}
\label{obsanddr}

\subsection{CXO Observations}
In our latest {\sl CXO} campaign B1259 was observed seven times (ObsIDs 16822, 16823, 18744, 19280, 20054, 19281, 20116) with the Advanced CCD Imaging Spectrometer (ACIS) between 2015 April and 2017 July.
Because observations 16823 and 18744, as well as 19281 and 20116, were carried out with very small time intervals between them, 4 and 3 days, respectively, each of these two pairs can be considered as a single observation, i.e., the binary was observed in 5 epochs. The pulsar's orbital position during each observation can be seen in Figure \ref{pic1}, while the specific observation dates, exposure times, days after periastron at time of observation, and true anomaly of the pulsar for each observation can be seen in Table \ref{tab1}. Similar to our previous observations reported in K+14 and P+15, we imaged B1259 on the front-illuminated ACIS-I3 chip in timed-exposure mode. The data were telemetered in the ``very faint'' format, and a 1/8 subarray was used to reduce the frame time to 0.4 s in order to lessen the effect of pile-up. The largest count rate from the binary was 0.05 counts per frame, which corresponds to a pile-up fraction of $<2\%$, and therefore can be neglected. 

The pipeline-produced Level 2 event files were used for all analyses. There were no episodes of anomalously high background rates occurring during any of these observations. In order to reduce the background contribution, all event files were filtered to only contain photon energies in the 0.5--8.0 keV energy range. The detector responses for spectral analyses were produced with the Chandra Interactive Analysis of Observations (CIAO, ver.\ 4.9) tools using the standard procedures and calibration database (CALDB ver.\ 4.7.5). The spectral fitting was done using XSPEC (ver.\ 12.9.1; \citealt{1996ASPC..101...17A}).

\subsection{HST Observations}
The Space Telescope Imaging Spectrograph (STIS) onboard the {\sl HST} has a wide-filter imaging mode, which can be used to take optical coronagraphic images with an occulting mask (50CORON). STIS has a $52''\times52''$ field of view, a 0\farcs05 pixel size, and operates in the 2000--10,300 ${\rm \AA}$ waveband. STIS observed B1259 on 2017 July 24 (MJD 57959) for four orbits ({\sl HST} program number 14932) shortly after the final {\sl CXO} observation (see Table 1). Each of the four observations were split into four shorter exposures of 648 s long (i.e., CR-SPLIT=4) to allow for the rejection of cosmic rays. The bright companion ($V\approx 10$ mag) star was placed on the WEDGEB2.5, which has an occultation width of 2\farcs5\footnote{See \url{http://www.stsci.edu/hst/stis/documents/handbooks/currentIHB/c12_special12.html}.}. We observed B1259 with four different roll angles: ORIENT=68\fdegm056, 71\fdegm056, 74\fdegm0559, and 77\fdegm0559, one per orbit. This is used to remove the bright diffraction spikes in the point spread function (PSF) wings of the companion star  while searching for extended optical emission from the clump (see Section \ref{hstiman}).

\section{Data Analysis}
\label{DA}

\subsection{X-ray Images}
\label{ximg}

\begin{figure*}
\centering
\includegraphics[trim={50 0 0 0},width=18.0cm]{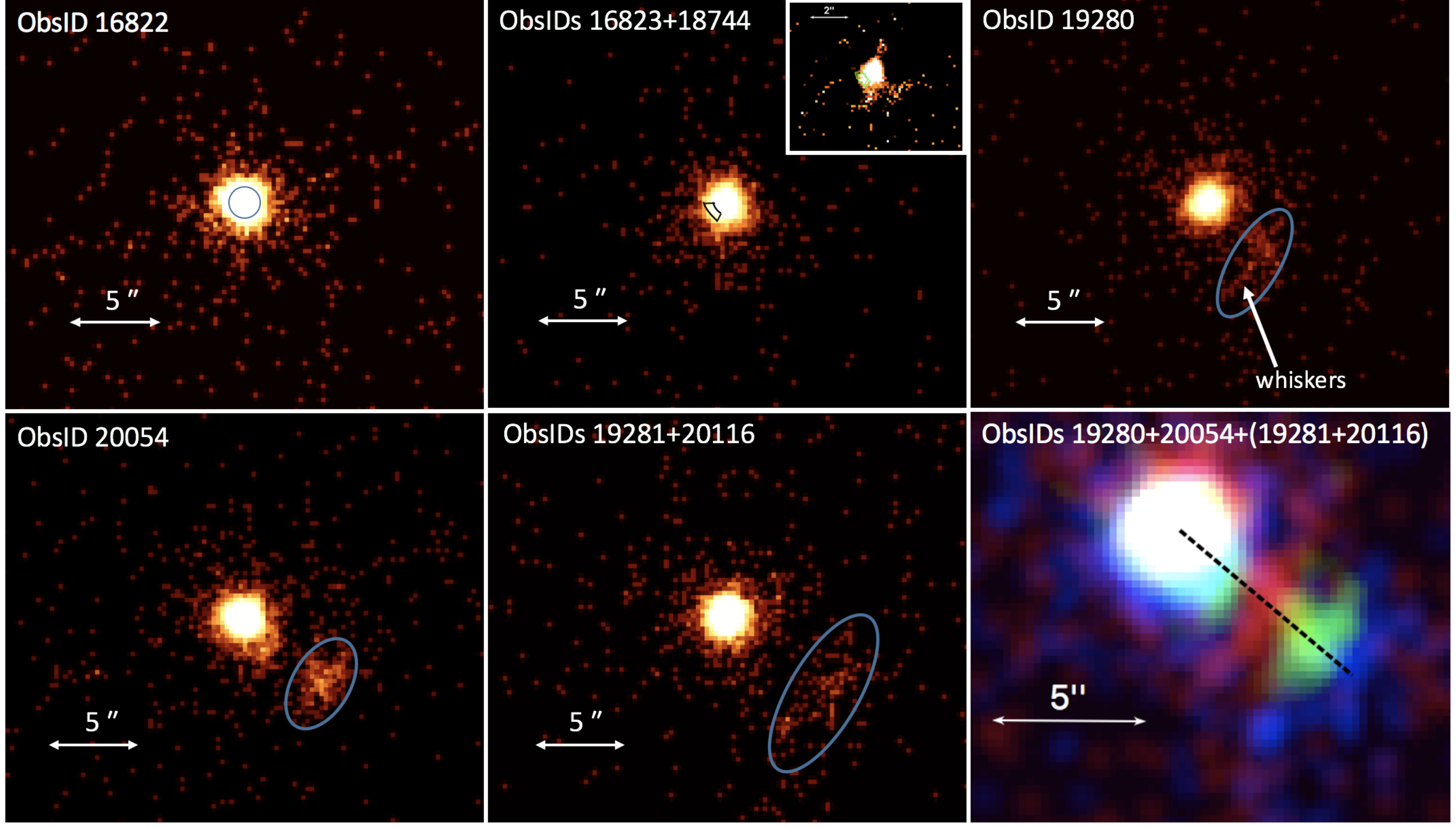}
\caption{ACIS 0.5-8 keV images from our 2014-2017 observing campaign are shown in the first five panels. Images from the last three observations are shown together, in different colors, in the bottom right panel to demonstrate the direction of motion and changes in morphology of the clump. The dashed black line connects the center of the binary to the centroids of the extended feature. The spectral extraction regions are shown in blue (an $r=1\farcs1$ circle for a representative extraction region for the binary core and ellipses for the extended feature). The inset in the top middle panel shows the {\tt arestore} results after 25 iterations (see Section \ref{ximg}). The green region in the inset shows the residual due to the unmodeled mirror asymmetry. 
\label{pic2}
}
\end{figure*}

 The sequence of five {\sl CXO} images obtained in the 2014-2017 binary cycle show a new X-ray emitting clump moving away from the binary and allows one to track the clump's motion up to a distance of $\sim7''$ (see Figure \ref{pic2}). In the last 3 observations, once the clump was clearly detached from the binary, we measured the positions and corresponding 3$\sigma$ uncertainties of the binary and the clump using CIAO's {\tt celldetect} tool and then calculated their separation.

To analyze the image structure more carefully, particularly in the cases where the presence of extended emission was not obvious (like in the first two observations), we used the Lucy-Richardson deconvolution algorithm (\citealt{1974AJ.....79..745L}; \citealt{1972JOSA...62...55R}). This algorithm deconvolves the PSF from the brightness distribution intrinsic to the source and is implemented by the CIAO tool {\tt arestore}\footnote{See \url{http://cxc.harvard.edu/ciao/ahelp/arestore.html}.}. An accurate PSF must be provided to {\tt arestore}, so the PSF of the observations are modeled using the Chandra Ray Tracer (ChaRT) and projected onto the ACIS-I detector plane using the MARX package (\citealt{2012SPIE.8443E..1AD}; \citealt{2003ASPC..295..477C}). The best-fit spectral model for the emission from the binary (see Table \ref{tab1}) was used in the PSF simulation.

The inspection of  the ObsID 16822 does not show any credible extended emission, so we can only place an upper limit of $\approx1\farcs5$ on the distance of any object that is detached form the binary (see Figure \ref{pic4}) assuming that the emission is as bright or brighter than that seen in the next observation.  

The first evidence of extended emission can be seen in the merged images of ObsIDs 16823 and 18744 (618 and 622 days after the 2014 periastron passage, respectively). It appears as an extended feature around $1\farcs7\pm0\farcs5$
south-southwest from the binary\footnote{CIAO's {\tt celldetect} could not detect the extended feature because it is too close to the binary, so we calculate the distance to the feature using the peak in the counts distribution for the binary and clump. The uncertainty is equal to one-half of the width of the feature.} in the merged deconvolved image (see the inset in the second panel of Figure \ref{pic2}). We 
examined the location of the mirror asymmetry\footnote{See \url{http://cxc.harvard.edu/ciao/caveats/psf_artifact.html} for more details.} for this observation and found it to be imaged southeast of the binary, away from the extended emission (see the green regions in the inset in the second panel of Figure \ref{pic2}).

The next observation, taken 978 days after periastron passage (ObsID 19280), clearly reveals a clump of extended emission, which is separated from the binary. There appears to be some sub-structure to this extended emission that was not seen in the previous 
observations (see Figure 2 in P+15). Most notably, there is a linear structure that extends perpendicular to the direction of the clump's apparent motion, which we nickname the ``whiskers'' (shown by a white arrow in Figure \ref{pic2}). This observation shows that the clump has traveled $3\farcs8\pm0\farcs7$ away from the binary.

Once the clump had been detected, a {\sl CXO} DDT observation (ObsID 20054) was taken 1086 days after periastron passage to better constrain the clump's motion and to monitor its flux evolution. Surprisingly, the new observation showed a clear brightening of the clump and a change in morphology, in which it takes on a bow-shock like shape. The clump was $5\farcs5\pm0\farcs6$ away from the binary. Further, the image shows a second clump of emission detaching from the binary in the same direction as the first clump. It lies at a distance of 2\farcs2$\pm0\farcs5$ from the binary (see Figure \ref{pic13} and Section \ref{discuss}).

The final observations (ObsIDS 19281+20116) of the campaign were carried out 1173 and 1176 days after periastron passage, respectively. These observations revealed the apparent dispersing and dimming of the clump. In the latest observation, the clump appears to be 6\farcs7$\pm1\farcs2$ away from the binary. The second clump detected in ObsID 20054 is not detected in this observation or in its deconvolved image.

Images from the three last observations are combined, with different colors, in the sixth panel of Figure \ref{pic2}. We see that the clump was moving along a straight line at a position angle of 230$^{\circ}$ (north through east), similar to the
215$^{\circ}$ angle measured in the previous binary cycle (P+15). The excellent angular resolution of {\sl CXO} allowed us to observe changes in the clump's morphology and brightness, which are much more drastic than those of the clump observed in the previous cycle (P+15).

\subsection{Projected Motion of the Extended Emission}
\label{mot}

We fit a linear model to the data, $r(t)=\mu(t-t_0)+r_0$, to determine the projected velocity of the clump. We chose $t_0=964.25$ days as the reference time, which is the mean value of the time intervals that have passed between periastron passage and each of the observations. The best-fit parameters are $r_0=4\farcs4\pm0\farcs3$ and $\mu=0\farcs008\pm0\farcs001$ day$^{-1}$ or 3\farcs0$\pm$0\farcs5 year$^{-1}$, which corresponds to a projected velocity $v_{\bot}=(0.12\pm0.02)c$ at a distance  $d=2.6$ kpc. However, the launch time (defined as $r(t_{\rm launch})=0$) is $t_{\rm launch}=420^{+78}_{-103}$ days for this fit. At this launch time, the pulsar is far away ($>9$ AU) from its stellar companion (see ObsID 16822 in Figure \ref{pic1}) and has long since passed through periastron and its companion's disk. Therefore, we consider this to be an  unlikely scenario because it is difficult to imagine a mechanism that could launch the clump with such a large initial velocity with minimal interaction between the pulsar and companion's winds. In the previous observations, the launch time was consistent with periastron passage and the  disk crossings (P+15; see Figure \ref{pic4} below).

If we assume that the launch time of the clump should occur near periastron passage, we can restrict our models to uphold this criterion. We first fit a linear model $r(t)=\mu t$, assuming  $r_0=0$. The best fit proper motion for this model is $\mu=0\farcs0044\pm0\farcs0006$ day$^{-1}$ or $1\farcs6\pm0\farcs2$ year$^{-1}$, corresponding to a projected velocity  $v_{\bot}=(0.066\pm0.009)c$ with a $\chi^{2}=7.8$ for 3 degrees of freedom. Due to the poor fit of this model, we also fit a model with an acceleration term: 
 $r(t)=r_0+v_{0,\perp}(t-t_0)+a_{\perp}(t-t_0)^2/2$, with $r_0=0$, $v_{0,\perp}=0$, and $t_0=0$, i.e., $r(t)=a_\perp t^2/2$. The best-fit acceleration is $a_\perp=(9.0\pm0.4)\times10^{-6}$ arcseconds day$^{-2}$ or 47$\pm2$ cm s$^{-2}$ or (14,800$\pm600$) (km/s) yr$^{-1}$ with $\chi^{2}=0.80$ for 3 degrees of freedom (see Figure \ref{pic5}).

\begin{figure}
\centering
\includegraphics[trim={0 0 0 0},width=8.5cm]{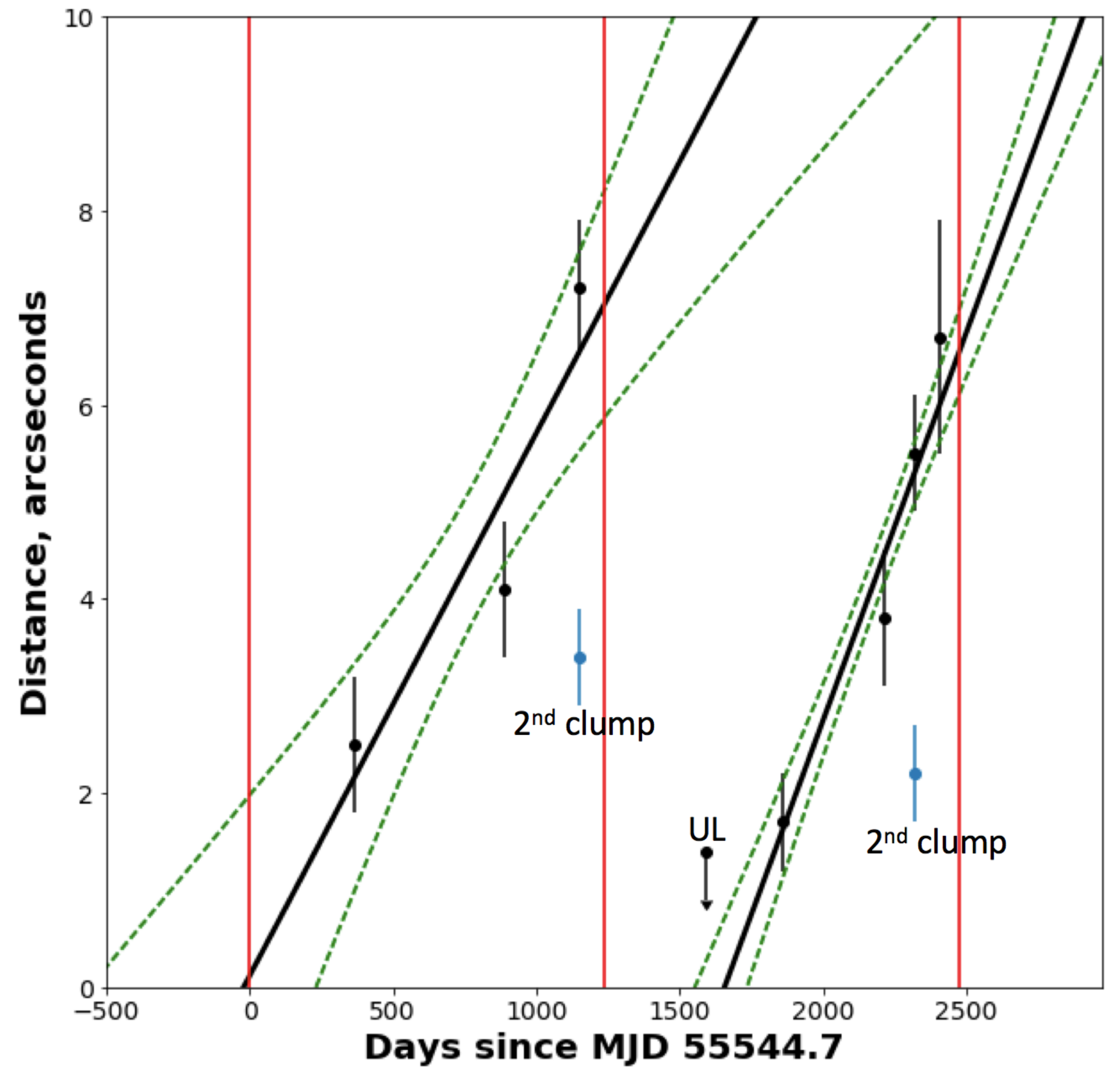}
\caption{Separation between the extended feature and unresolved binary as a function of time since the 2010 periastron passage.  Vertical red lines show the dates of periastron passages that took place on 2010 Dec 14 (MJD 55544), 2014 May 4 (MJD 56781) and 2017 Sept 22 (MJD 58018). The lines of best-fit and 1$\sigma$ upper and lower bounds are shown for the data reported in P+15 and the new data from the 2014-2017 orbital cycle. The point labeled ``UL'' is an upper limit on the distance for ObsID 16822. The separations of the second clumps, seen in ObsIDs 16553+16583 (in the 2010-2014 cycle) and 20054, from the binary are shown as blue points.
\label{pic4}
}
\end{figure}

\begin{figure*}
\centering
\includegraphics[trim={0 0 0 0},width=18.0cm]{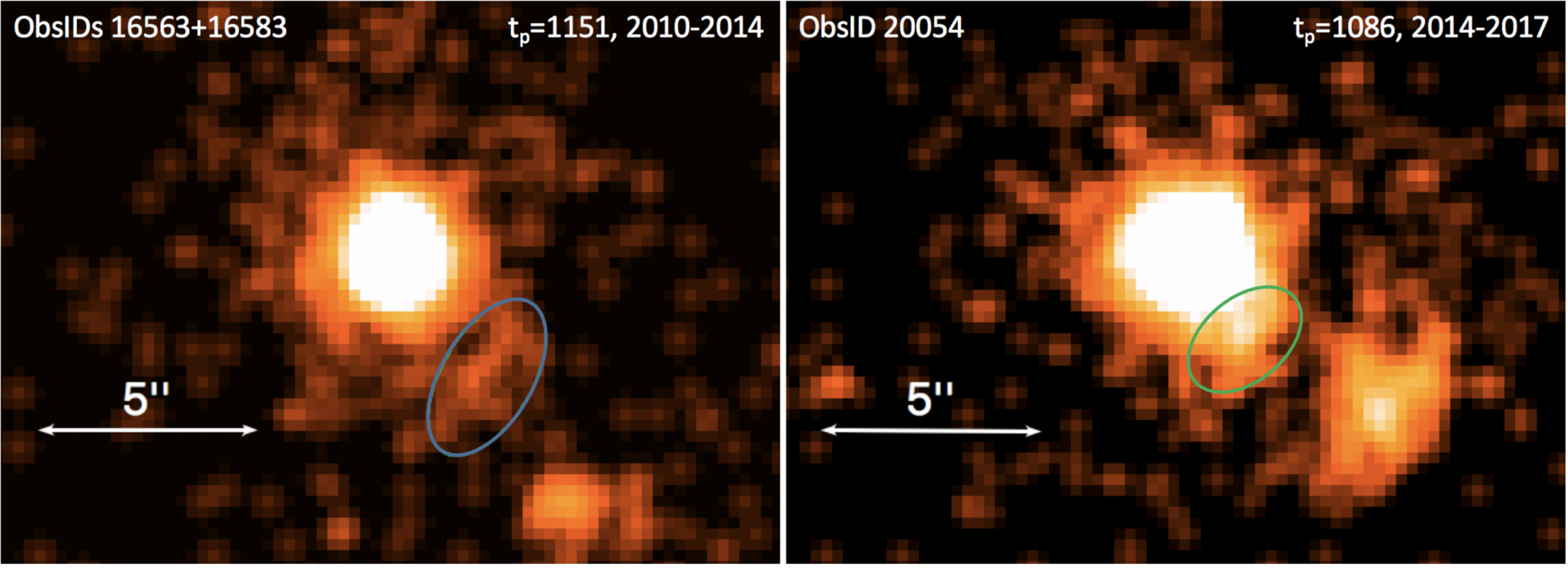}
\caption{ACIS 0.5-8 keV images from our 2010-2014 (obsIDs 16563+16583, left) and 2014-2017 (obsID 20054, right) observing campaigns. The images are smoothed using a Gaussian kernel with a size of $2''$. The observations,  taken 1151 and 1055 days after periastron passage, show evidence of a second clump (shown by  ellipses) being launched from the binary. 
\label{pic13}
}
\end{figure*}

\begin{figure}
\centering
\includegraphics[trim={0 0 0 0},width=8.5cm]{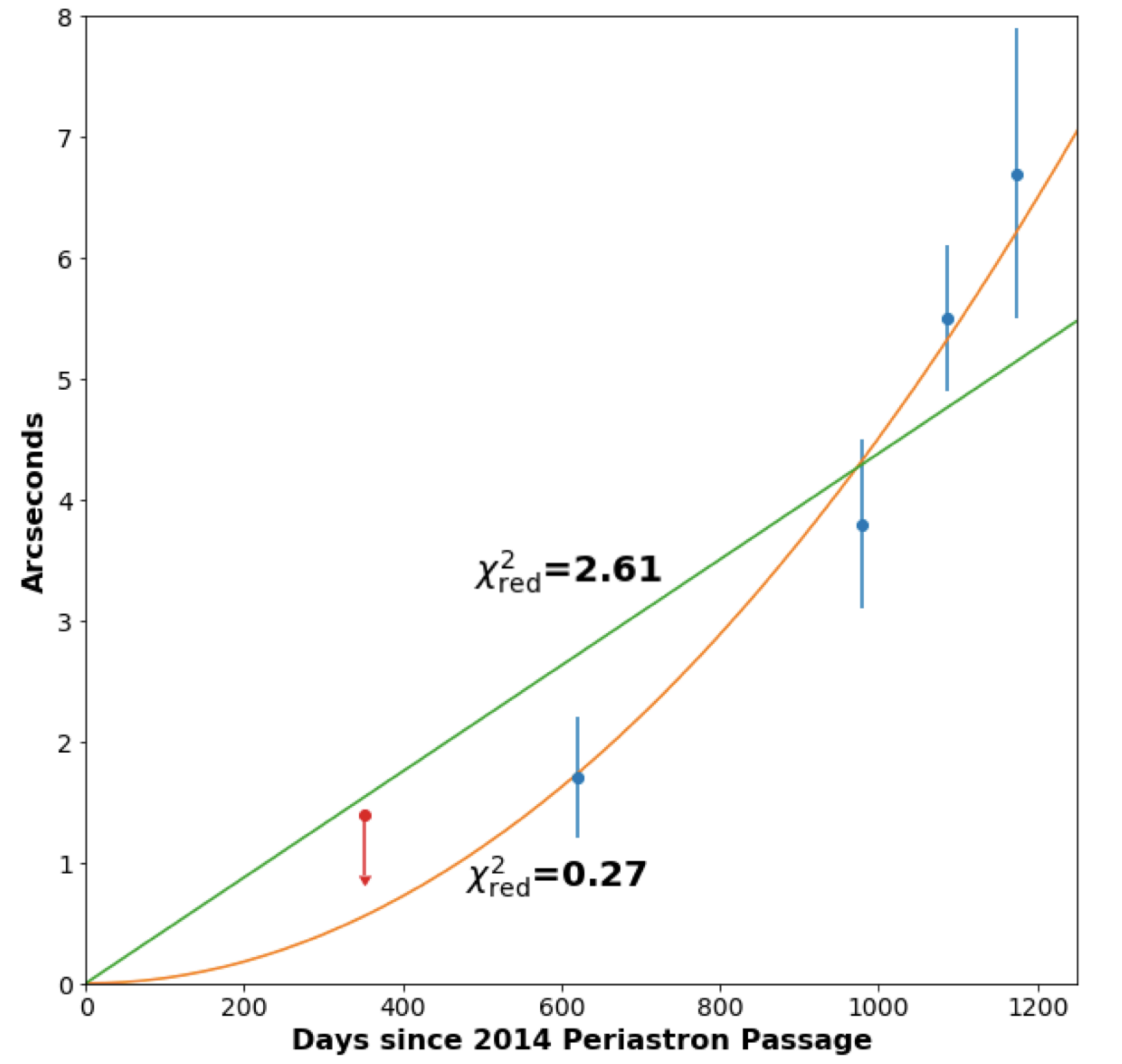}
\caption{Best-fit constant velocity ($v_\perp =0.066 c$; green) and 
constant acceleration ($a_\perp=47$ cm s$^{-2}$; yellow)
 models for the separation of the clump from the binary as a function of time. The models assume the clump is launched at periastron (see Section \ref{mot}). The reduced chi-squared ($\chi_{\rm red}^2$) for each model is shown. 
 \label{pic5}
}
\end{figure}

\subsection{HST image analysis}
\label{hstiman}

\begin{figure*}
\centering
\includegraphics[width=18.0cm]{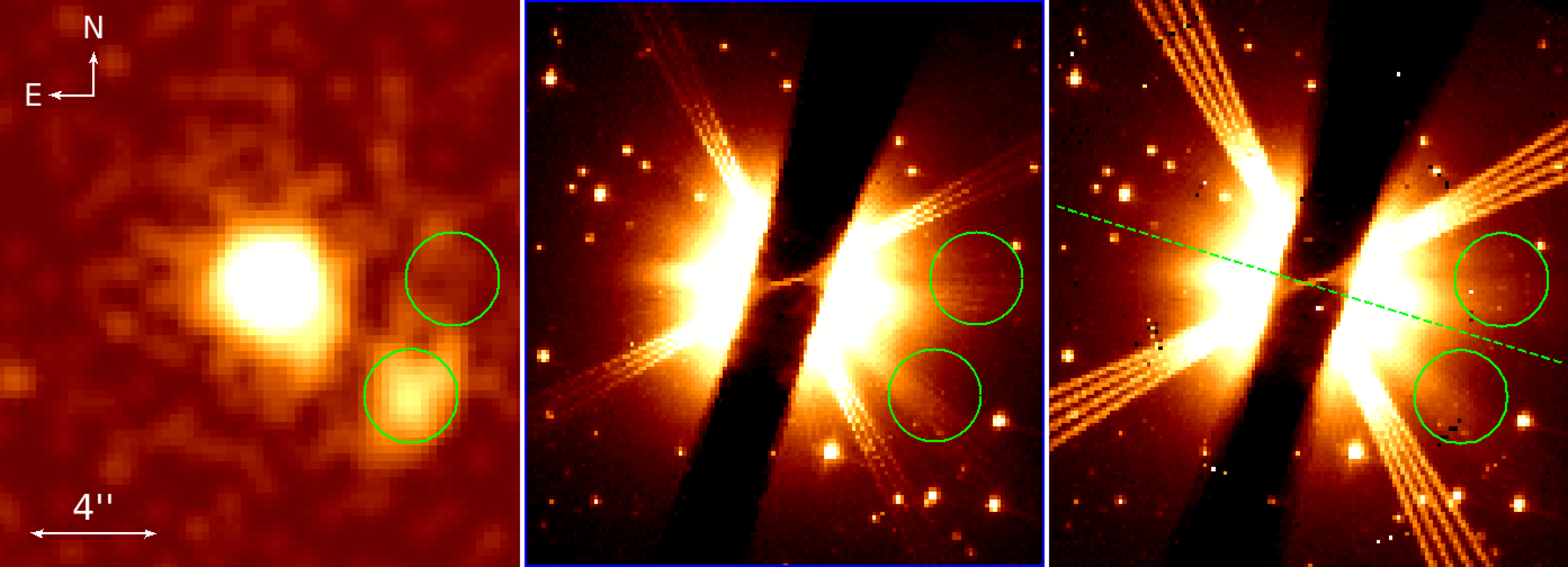}
\caption{ {\sl Chandra} ACIS (ObsID 20054; left)  and  {\sl HST} STIS CCD (right and middle)  images of the binary and extended X-ray clump. The min/max
filtered image is shown in the middle panel,  and the summed image is shown in the right panel. The solid circles ($r=1\farcs45$) show the source and background extraction regions. The dashed line 
is the line of mirror symmetry, about which the source and background regions are symmetric.
\label{hst}
}
\end{figure*}

The {\sl HST} STIS CCD images of the clump region are shown in Figure 
\ref{hst}  (middle and right panels)  together with the nearly 
contemporaneous {\sl Chandra} ACIS image (left panel). In order to 
eliminate the small scale  artifacts (e.g., pixels with abnormally large values seen in the summed image), and to reduce the 
impact of  the bright diffraction spikes associated with the PSF, we have produced both the min/max filtered image (middle), and summed image (right).   For the min/max filtered  image,  
we retained two out of the four values (corresponding 
to the four observations) for each pixel by throwing out the minimum and maximum values. Note that, although the filtering somewhat improved the image, the wings of the  bright star's PSF were unable to be subtracted at the location of the X-ray clump due to the asymmetric shape of the PSF,  which has multiple fainter diffraction 
spikes (in addition to the brightest four spikes).
 We also note that the PSF is 
asymmetric with respect the line drawn through the middle of the 
occulting wedge in the image, possibly because the center of the star was 
slightly offset from this line. This latter issue prompted us to 
choose the source and background regions on the same side of the 
occulting wedge, where both regions encompass a similar PSF structure (see Figure \ref{hst}). 

In order to estimate the impact of 
small-scale non-uniformities caused by the remaining fainter narrow 
diffraction spikes, we performed multiple (10 for each region)  source 
and background measurements by  shifting the source and background 
regions (shown in Figure \ref{hst}) in random directions  by 35$-$70 mas.
From these measurements, we found $(89.7\pm2.1)\times 10^3$ and $(86.3\pm2.5)\times10^3$ counts
in the source 
and the background regions,
respectively, in the 5176 s exposure. For 
comparison, we found $(30.8\pm1.3)\times10^3$ counts (within the same exposure time) in the same sized aperture placed 
at randomly chosen locations far away from any bright stars.
The net number of 
counts in the source region  is 
$n_s=3.4\times10^{3}$ and its standard deviation is 
$\sigma_s=3.3\times10^3$. Therefore, we conclude that the clump is not 
detected in our  {\sl HST} observations.  
The $3\sigma$ upper limit on the source count rate of 
1.9 counts s$^{-1}$, 
translates 
into an 
unabsorbed (dereddened) spectral flux of 280 nJy (or an observed spectral flux of 29 nJy) at $\nu=5.2\times 10^{14}$ Hz ($\lambda = 5769\,\AA$).
These estimates were obtained using the {\tt Synphot} package, assuming
a power-law spectrum, $F_\nu\propto\nu^{-0.45}$, with the same slope as
the clump's X-ray spectrum (see Sec. \ref{CXO_spec_an}) and a color excess of $E(B-V)=0.85$ 
\citep{2011ApJ...732L..11N}.

\begin{figure*}
\centering
\includegraphics[width=18.0cm]{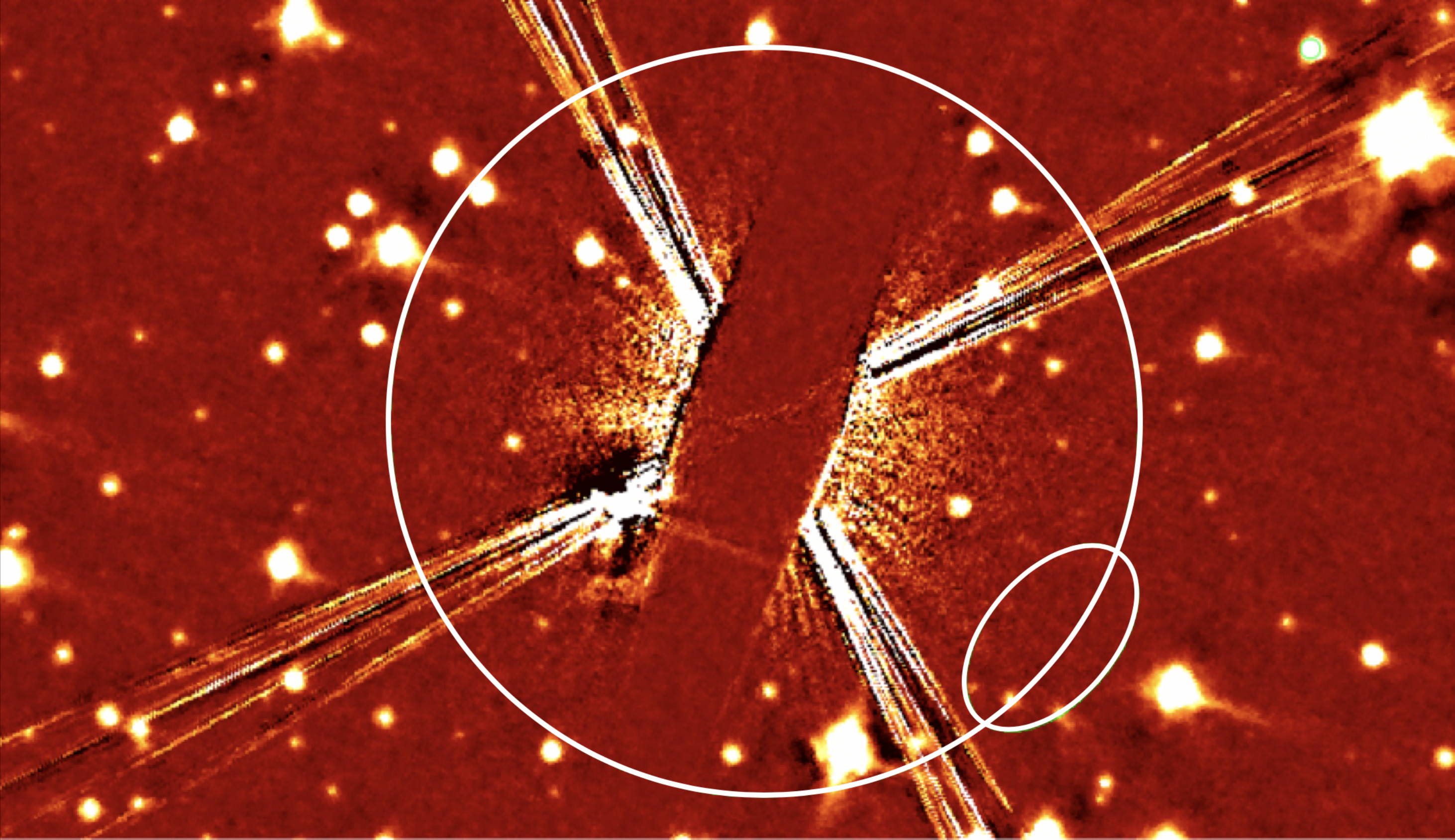}
\caption{ The combined {\sl HST} STIS/CCD image  of the LS 2883 vicinity with the star being subtracted. No extended emission is  seen at the location of the X-ray clump (shown by the ellipse). The radius of the larger circle centered on the star is $6''$. North is up, East is to the left.
\label{hst2}
}
\end{figure*}

We have also produced an image where we have subtracted the emission of the bright star (see Figure \ref{hst2}) following the procedures typically used for exoplanet detections (see e.g., \citealt{2013ApJ...775...56K}). Here we briefly describe the procedure. We started with the pipeline processed  (flat-fielded, bias- and dark-subtracted, CR-rejected) images. As a first step, we applied a median spatial filter to each of the 4 images using a  $2\times2$ pixel (one pixel is $0.05''$) box to remove any remaining pixel-scale artifacts (e.g., noisy, bad, flaring pixels). In the next step, we produced a median-combined image out of the 4 images registered to the telescope's reference frame (RF).  Since the telescope was rotated by $3^{\circ}$ per observation, this step removes most of the stars in the image and produces a clean image of the PSF from the bright star occulted by the coronagraph. The median image of the PSF is then rotated to coincide with the telescope orientation during each individual observation and is then subtracted from each individual image. Finally, the resulting images are added together, producing  a single deep image with the bright star  subtracted out (see Figure \ref{hst2}).  There are no noticeable extended structures within the area with the size of the X-ray clump (the white ellipse in Figure \ref{hst2}) suggesting that its emission is not detected by {\sl HST}. However,  we do not use this image for the upper limit estimation because the azimuthal extent of the clump in the X-ray images ($\approx2''$) significantly exceeds the $\approx0.3''$ shifts (due to the telescope's  rotation) between the individual images at the $6''$ distance between the star and clump. Therefore, the clump's emission is being partly subtracted from itself while making the median-combined image of the central star. This method could be more beneficial in the future if the clump is farther from the star, is more compact, or the angle increments of the telescope's rotation are larger.

\subsection{CXO Spectral Analysis}
\label{CXO_spec_an}

The spectra from both the binary core and the clump were extracted for each observation, but those obtained from either ObsIDs 16823+18744 or 19281+20116 were fit simultaneously. Spectra from the binary were taken from an $r=1\farcs1$ radius circle placed at the binary center and then grouped to contain 25 spectral counts per bin. An absorbed power-law model, using the XSPEC {\tt tbabs} absorption model, was used to fit the spectra from both the binary and the clump. The fit parameters obtained from each observation and subsequent fit can be seen in Table \ref{tab1}.

\begin{figure*}
\centering
\includegraphics[trim={0 0 0 0},width=18.0cm]{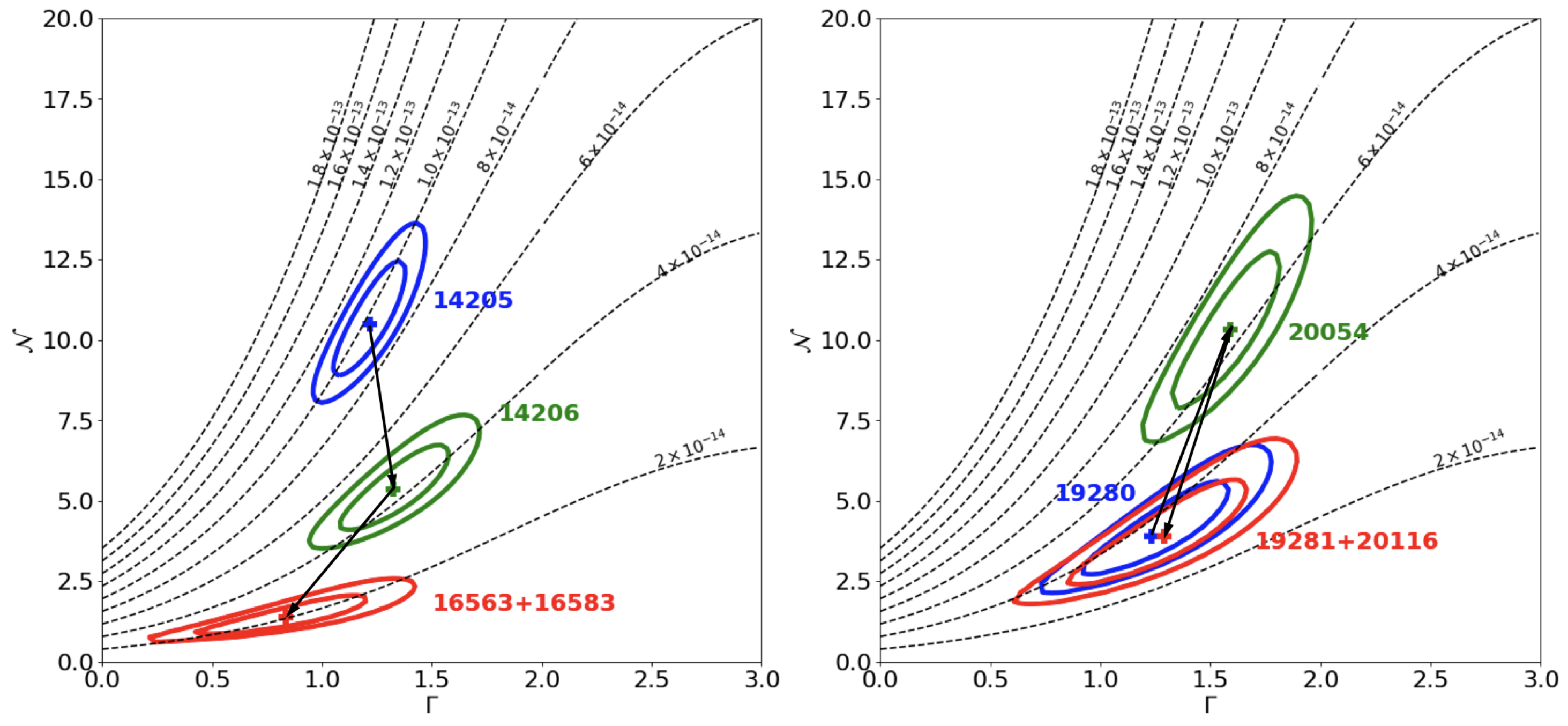}
\caption{Confidence contours (90\% and 99\%) for the ACIS spectra of the extended feature in the $\Gamma-\mathcal{N}$ plane for the clump detected during the 2010-2014 (left) and 2014-2017 (right) orbital cycles. The extraction region definitions can be found in Figure \ref{pic2}, P+15, and K+14. The color and arrows show the time order of the observations (blue is the observation taken nearest to periastron passage, then green, then red).
\label{pic3}
}
\end{figure*}

The clump was not seen in ObsID 16822, and it was too close to the binary to extract an uncontaminated spectrum in ObsIDs 16823+18744, so they have been omitted. To fit the spectra of the clump we used C-statistics\footnote{See \url{https://heasarc.gsfc.nasa.gov/xanadu/xspec/manual/XSappendixStatistics.html}.}  \citep{1979ApJ...228..939C}. The hydrogen absorption column density ($N_{\rm H}$) was fixed for each spectral fit to the value obtained from the fit of the binary core. All best-fit parameters are provided in Table \ref{tab1}. Figure \ref{pic3} shows the  1$\sigma$ and 2$\sigma$ confidence contours in the normalization (${\cal N}$) versus photon index ($\Gamma$) plane for all observations where it was possible to extract a spectrum. Interestingly, the flux of the clump increases by a factor of $\sim1.5$ in between ObsIDs 19280 and 20054, and then dims in the final observation (ObsIDs 19281+20116). This implies brightening on timescales of $\lesssim 100$ days, which was not seen in the previous observations (see Figure \ref{pic10}, K+14, P+15). 

The photon index of the clump remains consistent across all observations and is also consistent with the previous observations (K+14, P+15). Therefore, we fit all extracted spectra from the 2014-2017 orbital cycle simultaneously. The hydrogen absorption column density was frozen to the mean value ($N_{\rm H}=3.1\times10^{21}$ cm$^{-2}$) of the observations for which spectra of the clump were extracted. The normalizations of each spectrum were left free, since the clump changes its brightness. This fit gives us a more constrained photon index ($\Gamma=1.45\pm0.11$), assuming that the photon index does remain constant across all observations. The unabsorbed 0.5-8 keV luminosity of the clump is $\sim(3-6)\times10^{31}$ erg s$^{-1}$ at $d=2.6$ kpc, which is a few percent of the binary's X-ray luminosity.

\begin{table*}
\caption{Spectral fit parameters for the core and extended emission in the five \cxo ACIS observations} 
\label{tab1}
\begin{center}
\renewcommand{\tabcolsep}{0.11cm}
\begin{tabular}{lcccccccccccc}
\tableline 
ObsID	&  MJD &	$\theta$\tablenotemark{a} 	&	$\Delta t$\tablenotemark{b}	&	Exp.\tablenotemark{c}	&	Cts\tablenotemark{d}	& 
$F_{\rm obs}$\tablenotemark{e}	&	$F_{\rm corr}$\tablenotemark{f}	&	$N_H$	&	$\Gamma$	&	$\mathcal{N}$\tablenotemark{g}	&
$\mathcal{A}$\tablenotemark{h} 	&	$\chi^2$/d.o.f.\tablenotemark{i} \\
&  &	deg	&	days	&	ks	&	&	$10^{-14}$ cgs	&	$10^{-14}$ cgs	&	$10^{21}$ cm$^{-2}$	&	&
$10^{-4}$	&	arcsec$^2$	& \\
\tableline 
16822 & 57133 & 169 & 352 & 58.3 & 6176 &  148$^{+2}_{-3}$ & 185$^{+2}_{-3}$ & 3.4(3) & 1.57(4) & 2.9(2) & 3.8 & 178.0/176\\
  & & & & & ... & ... & ... & ... & ... & ... & ... & ... \\
  16823+18744\tablenotemark{j} & 57401 & 180 & 620 &32.4$+$28.6 & 4315 & 101$^{+1}_{-2}$ & 123$^{+1}_{-2}$ & 2.7(4) & 1.59(5)  & 2.0(1) & 3.8 & 132.4/145\\
  & & & & & ... & ... & ... & ... & ... & ... & ... & ... \\
   19280 & 57759 & 198 & 978 & 55.4 & 5191 & 124(3) & 172(2) & 3.4(3) & 1.94(5) & 3.7(2) & 3.8 & 161.6/148\\
  & & & & & 109 & 3.0$^{+0.3}_{-0.5}$ & 3.5$^{+0.3}_{-0.5}$ & 3.4$^\ast$ & 1.2(2) & 0.040$^{+0.009}_{-0.008}$ & 14.4 & 67.2/93 (C) \\
    20054 & 57867 & 208 & 1086 & 59.0 & 7382 & 181(3) & 232(3) & 3.2(3) & 1.70(4) & 4.1(2) & 3.8 & 247.5/193\\
  & & & & & 203 & 5.2$^{+0.6}_{-0.5}$ & 6.5$^{+0.6}_{-0.5}$ & 3.2$^\ast$ & 1.6(2) & 0.10$^{+0.02}_{-0.01}$ & 13.3 & 120.7/150 (C) \\
  19281+20116\tablenotemark{j} & 57956 & 227 & 1175 & 35.5+19 & 6267 & 178(3)   & 213(3)  & 2.6(3)  & 1.52(4)  &  3.2(2) & 3.8 & 206.7/200 \\
 & & & & & 106 & 2.9$^{+0.4}_{-0.5}$ & 3.3$^{+0.4}_{-0.5}$ & 2.6$^\ast$ & 1.3(3) & 0.038$^{+0.010}_{-0.009}$ & 29.8 & 88.6/94 (C) \\
\tableline 
\end{tabular} 
\end{center}
\tablecomments{For each ObsID the upper and the lower rows correspond to the binary core and extended emission, respectively. The $1\sigma$ uncertainties of the last significant digit are shown in parentheses. Fluxes and counts are in the 0.5--8\,keV range. An asterisk indicates that the hydrogen absorption column density was fixed to the value of the corresponding point source fit. For observations that were fit simultaneously, we provide the mean values of the date, true anomaly, and days after periastron. }
\tablenotetext{\textnormal{a}}{True anomaly counted from periastron.}
\tablenotetext{\textnormal{b}}{Days since 2014 periastron passage (MJD 56781).}
\tablenotetext{\textnormal{c}}{Exposure corrected for deadtime.}
\tablenotetext{\textnormal{d}}{Total (gross) counts.}
\tablenotetext{\textnormal{e}}{Observed flux.}
\tablenotetext{\textnormal{f}}{Extinction-corrected flux calculated using {\sl CXO} WebPIMMS \url{http://cxc.harvard.edu/toolkit/pimms.jsp}.}
\tablenotetext{\textnormal{g}}{Normalization in photons\,s$^{-1}$\,cm$^{-2}$\,keV$^{-1}$ at 1\,keV.}
\tablenotetext{\textnormal{h}}{Area of the extraction region.}
\tablenotetext{\textnormal{i}}{If the data were fit by minimizing C-stat instead of $\chi^2$, then the C-stat/d.o.f. is reported and denoted by (C). }
\tablenotetext{\textnormal{j}}{The spectra from the two observations were fit simultaneously.}
\end{table*}

\section{DISCUSSION}
\label{discuss}

The appearance of the clump after periastron passage
led P+15 to suggest that the clump, composed of fragments
of the massive star's decretion disk, is launched when the pulsar interacts with the disk.
  However, if the clump is moving in the stellar wind of the high-mass companion, it should experience a large drag force 
  with a characteristic deceleration time
  of $\sim10$ s  (see K+14). 
  Yet, the clumps 
observed in both orbital cycles show no signs of deceleration on timescales of several years.  In order to overcome this difficulty,  P+15 suggested that the
     clump is moving  in (and  being propelled by) the fast-streaming pulsar wind.
In the scenario proposed by P+15,  the clump is moving along the periastron-to-apastron direction in the unshocked PW,   which is dynamically dominant, i.e., $\eta >1$, where $\eta\equiv\dot{E}/(\dot{M}_*v_w c)$ is the ratio of pulsar to stellar wind momentum fluxes, $\dot{M}_*$ is the companion's mass-loss rate due to its isotropic wind, and $v_w$ is the companion wind's velocity. Alternatively, the simulations of \cite{2016MNRAS.456L..64B} have shown that the accelerated motion of the clump in the PW, which is shocked near (or even within) the binary, is still possible for $\eta<1$, i.e., when the stellar wind is dynamically dominant. 
This is because a  ``channel''  in the stellar
wind is carved out by the PW  streaming
 in the apastron direction (because the pulsar
spends most of the time near apastron).

Depending on the scenario (i.e., $\eta>1$ versus $\eta<1$) the fraction of the PW that interacts with the clump will vary. 
 If the PW is isotropic and dynamically dominant ($\eta>1$), then only a fraction $\xi_{\Omega}=(r_{\rm cl}/2r)^{2}\approx0.04$ of the PW is intercepted by the clump, where $r_{\rm cl}$ is the radial size of the clump, and $r$ is the distance between the clump and the binary. On the other hand, if the PW is confined to a channel, and the clump fills the entire cross section of the channel, then all of the PW interacts with the clump (i.e., $\xi_\Omega=1$). In reality, this factor $\xi_{\Omega}$, which accounts for the geometry of the PW flow, is likely to be in between these two limits so we carry it through our estimates.

\subsection{Radiation mechanism}
\label{rad_mech}

Once the dense clump is entrained into the very fast  pulsar wind,  a shock is expected to form at the PW-clump interface,
 regardless of the value of $\eta$ or whether the PW is shocked or unshocked by the time it reaches the clump\footnote{
However, the details of this interaction (e.g., particle acceleration at the shock) may be different for large and small $\eta$.}.
  If the X-ray emission of the clump is indeed powered by the PW, then the X-ray luminosity can be expressed as $L_{X,cl}=\xi_X\xi_{\Omega}\dot{E}$,
 where $\xi_X$ is the X-ray efficiency. 
 The observed luminosities of the clump, $L_{X,cl}\approx(3$--$6)\times10^{31}d_{2.6}^{2}$ erg s$^{-1}$, where $d= 2.6 d_{2.6}$ kpc is the distance to the binary, correspond to $\xi_X\xi_\Omega\sim(4-7)\times10^{-5}$. 
This implies an X-ray efficiency, $\xi_X \sim 10^{-4}$--$ 10^{-3}$, of the same order of magnitude  as those of synchrotron PWNe observed around isolated young pulsars \citep{2008AIPC..983..171K}.
  
  Another source of energy that could, in principle, power the clump's X-ray emission 
 is  the radiation field of the massive star.  The star's luminosity, $L_*=2.4\times10^{38}$ erg s$^{-1}$,
  is a factor of 300 larger than the pulsar's $\dot{E}$. 
   The energy density of the radiation field, $u_{\rm rad}=L_*/(4\pi c r^2)=6.4\times10^{-8} r_{17}^{-2}$   erg cm$^{-3}$ 
exceeds the magnetic field energy density, $u_B=B^2/8\pi=4.0\times10^{-8}(B/1~{\rm mG})^2$ erg cm$^{-3}$, for  $B<1.3 r_{17}^{-1}$ mG, where $r_{17}$ is the
clump's distance from the binary in units of $10^{17}$ cm.

  Below we discuss both  sources of power and conclude that, although the star produces more energy than the pulsar, it is unlikely to contribute significantly to the X-ray emission or the dynamics of the clump at the observed separations from the binary.

We first consider whether the scattering
of stellar photons off relativistic electrons  in  the clump could power the observed X-ray emission.
    Electrons with Lorentz factors exceeding $\gamma_{\rm KN}= m_e c^2/4\epsilon\sim10^4$, where $\epsilon\sim10$ eV is the typical energy of the most abundant UV photons produced by the star, will upscatter photons in the Klein-Nishina (KN) regime, while those with $\gamma<\gamma_{\rm KN}$ will scatter in the Thomson regime (see, e.g., \citealt{1970RvMP...42..237B}).  The broadband spectral energy distribution (SED) of the electrons is not known, but if the clump's X-ray emission is due to IC up-scattering of UV photons, it can only be due to up-scattering by electrons with $\gamma\sim10-100$ in the Thomson regime. Electrons with larger $\gamma$ will produce\footnote{Indeed, if IC upscattering  proceeds in the Thomson regime, then the UV photon energy is boosted by $\gamma^2$ (but remains a small fraction of $\gamma m_e c^2$), while if it occurs in the extreme KN regime (i.e., where $\gamma>\gamma_{\rm KN}$), then the typical energy gained by the photon in a single collision  is a sizable fraction of   $\gamma m_e c^2$ \citep{1970RvMP...42..237B}. } photons with  energies $\gtrsim 10$ keV.    
     
     Assuming that the dominant emission mechanism is IC, we can crudely estimate the lower limit on the number of IC emitting electrons as $N_e> L_X/P_{\rm IC}=1.4\times10^{50}(F_X/4\times 10^{-14}$ erg cm$^{-2}$ s$^{-1}$)$(\gamma/10)^{-2}r_{17}^{2}d_{2.6}^2$, where $P_{\rm IC}=(4/3)\sigma_Tc\gamma^2\beta^2u_{\rm rad}$ is the total IC power emitted by a single electron 
that upscatters stellar photons in the Thomson regime\footnote{Although the expression we used for $P_{\rm IC}$ is only valid for an isotropic photon field, it should be the same for the {\em average} scattered power (per electron) if the electrons come from an isotropic electron distribution (even if the radiation field is anisotropic as it happens to be in this case). }, and $F_X$ 
 is the observed X-ray flux of the clump in the 0.5--8 keV band. However, such a  large number of particles is problematic because the pulsar is expected to only supply $\dot{N}_e = 4 \pi^2 R_{\rm NS}^3 B_{\rm d} (ce)^{-1}P^{-2} \kappa_{\rm pair}=4\times 10^{32}\kappa_{\rm pair}$ electrons s$^{-1}$,  where $B_d=3.3\times10^{11}$ G is the dipolar field 
at the NS equator, $R_{NS}\sim10$  km is the NS radius, and $ \kappa_{\rm pair}\lesssim10^{5}$ is the pair cascade multiplicity \citep{2015ApJ...810..144T,2018arXiv180308924T}. Therefore, the maximum number of electrons injected by the pulsar since the launch of the clump  is $\dot{N}_e t \sim 3\times 10^{45}(\kappa_{\rm pair}/10^5) (t/1000~{\rm days})$.

Lastly, even if the clump contains the mass of the entire decretion disk ($m_{\rm disk}\sim10^{24}-10^{26}$ g;  \citealt{2014MNRAS.439..432C},  and  references  therein), the clump would contain $<6\times10^{49}$ electrons, and only a very small fraction of them could be accelerated to $\gamma \sim 10$--100. Therefore, we conclude (in agreement with \citealt{2011ApJ...730....2P} and P+15) that the IC mechanism is unlikely to be responsible  for the observed X-ray emission at distances of a few arcseconds away from the binary, and that the clump's X-ray emission is produced via synchrotron radiation.

\begin{figure}
\centering
\includegraphics[trim={0 0 0 0},width=8.5cm]{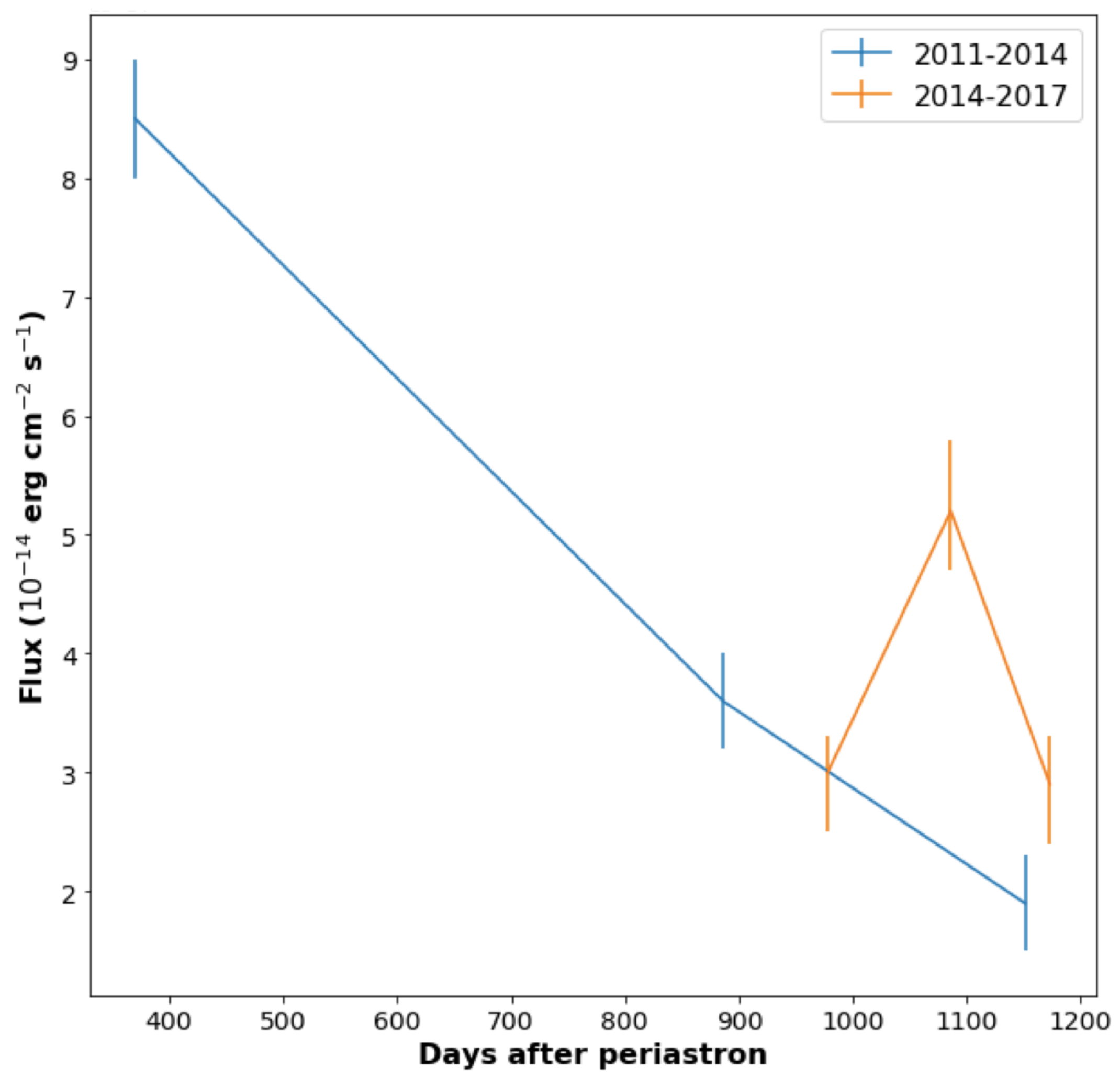}
\caption{Observed flux in units of $10^{-14}$ erg cm$^{-2}$ s$^{-1}$ versus time after periastron passage for the 2010-2014 (blue) and 2014-2017 (orange) orbital cycles.
\label{pic10}
}
\end{figure}

For relativistic particles accelerated within the clump (e.g., shocked  at the interface between the PW and the decretion disk material in the clump) the synchrotron emissivity depends on the magnetic field in the shock, the relativistic electron number density, and the Lorentz factors  of the emitting electrons. There are no significant spectral changes observed as the clump travels away from the binary. This, together with the relatively hard X-ray spectrum, suggests that synchrotron cooling does not have a noticeable impact  on the radiating electrons up to the distances where the clump is seen. This can happen either because of  continuous  re-acceleration, or if the synchrotron cooling time, $\tau_{\rm syn}=22 (B/100~\mu{\rm G})^{-3/2}(E_{\rm syn}/3~{\rm keV})^{-1/2}$ yrs, exceeds the dynamical time (i.e., the clump travel time\footnote{We assume that the clump is launched around the time of periastron passage.}) of $t\approx 3.4$ yrs. The latter condition implies $B\lesssim240(E_{\rm syn}/10~{\rm keV})^{-1/3}$~$\mu$G, assuming the magnetic field is constant or slowly varying over most of the dynamical time. On the other hand, the gyroradius of electrons must not exceed the, essentially unresolved, width of the ``whiskers'' in the ObsID~19280 image implying $B\gtrsim30(E_{\rm syn}/1~{\rm keV})^{-1/3}d_{2.6}^{-2/3}$~$\mu$G. 

These limits  comfortably  accommodate $B\approx80 k_m^{2/7} d_{2.6}^{-2/7}$~$\mu$G estimated by K+14 from the clump's surface brightness (where $k_m$ is the  unknown ratio  of  the  magnetic  field  energy  density  to the  energy  density  of  the  relativistic  particles).  Therefore, for a plausible magnetic field,  $q\equiv u_{\rm rad}/u_B\approx150 r_{17}^{-2}(B/100~\mu{\rm G})^{-2}\gg1$, which, according to Eqns.\ (8) and (9) from \cite{2005MNRAS.363..954M},  implies that synchrotron losses dominate IC losses for $\gamma> \gamma_s\simeq3.5\times10^{5}(B/100~\mu{\rm G})^{-4/3}r_{17}^{-4/3}$. The  Lorentz factors, $\gamma_1\sim 3\times10^{7}(B/100~\mu\rm{G})^{-1/2}$ and $\gamma_2\sim 10^{8}(B/100~\mu\rm{G})^{-1/2}$, of electrons that produce the 1--10 keV  synchrotron  photons are much larger than both  $\gamma_{\rm KN}\sim10^4$ and $\gamma_s$, implying that any IC scattering must occur in the KN regime, and that the synchrotron losses dominate the IC losses.

 The uncooled synchrotron interpretation of the observed X-ray spectrum also implies that the slope $p$ of the electron SED, $dN_e(\gamma)=K_e\gamma^{-p}d\gamma$ (where $\gamma_m<\gamma<\gamma_M$), is $p=2\Gamma-1\approx1.4-2.2$.  The photon index derived from fitting all three clump X-ray spectra simultaneously is $\Gamma=1.45\pm0.11$ corresponding to $p=1.85\pm0.22$. Although there are substantial uncertainties associated with the measurements of $\Gamma$ (see Table \ref{tab1}), the obtained $p$ values are below the value of $p\approx2.2$ typically expected for  Fermi-type shock acceleration \citep{2001MNRAS.328..393A}. This may suggest that an acceleration mechanism other than shock acceleration (e.g., magnetic reconnection) is responsible for the X-ray emission.

The number of electrons emitting synchroron radiation with a luminosity 
$L_{\rm syn}(E_1,E_2)\equiv L_X$ in the photon energy range ($E_1,E_2$)
 can be estimated as $N_e(\gamma_1,\gamma_2)\equiv N_{e,X} = L_X/P_{\rm syn}(\gamma_1,\gamma_2)$,
where $\gamma_i \sim (E_i/E_{\rm cyc})^{1/2}= 
2.9\times 10^7 (E_i/1\,{\rm keV})^{1/2} (B/100\,\mu{\rm G})^{-1/2}$ 
is the Lorentz factor of electrons that give the main contribution to radiation 
 at energy $E_i$, $E_{\rm cyc}=heB/(2\pi m_ec)$ is the cyclotron energy,
$P_{\rm syn}(\gamma_1,\gamma_2)=(4/3)\sigma_Tcu_B\langle \gamma^2\rangle$ is the mean synchrotron power per electron in the ($\gamma_1$, $\gamma_2$) range.
For the assumed power-law SED, the mean square of the Lorentz factor is
$\langle\gamma^2\rangle = C_{3-p}(\gamma_1,\gamma_2)/C_{1-p}(\gamma_1,\gamma_2)$, where $C_q(\gamma_1,\gamma_2) = (\gamma_2^q -\gamma_1^q)/q$. For instance, for $E_1=0.5$ keV, $E_2=8$ keV, and $p=1.85$, we have
$\langle\gamma^2\rangle \approx 1.7\times 10^{15}(B/100\,\mu{\rm G})^{-1}$, and $N_{e,X}\sim 8.2\times 10^{38} (F_{\rm 0.5-8\,keV}/4\times 10^{-14}\,{\rm erg\,cm^{-2}\,s^{-1}}) (B/100\,\mu{\rm G})^{-1} d_{2.6}^2 $. The number of electrons in this case is substantially smaller than the $>10^{50}$ required in the IC case (see above).

 \begin{figure}
\centering
\includegraphics[trim={0 0 0 0},width=8.6cm]{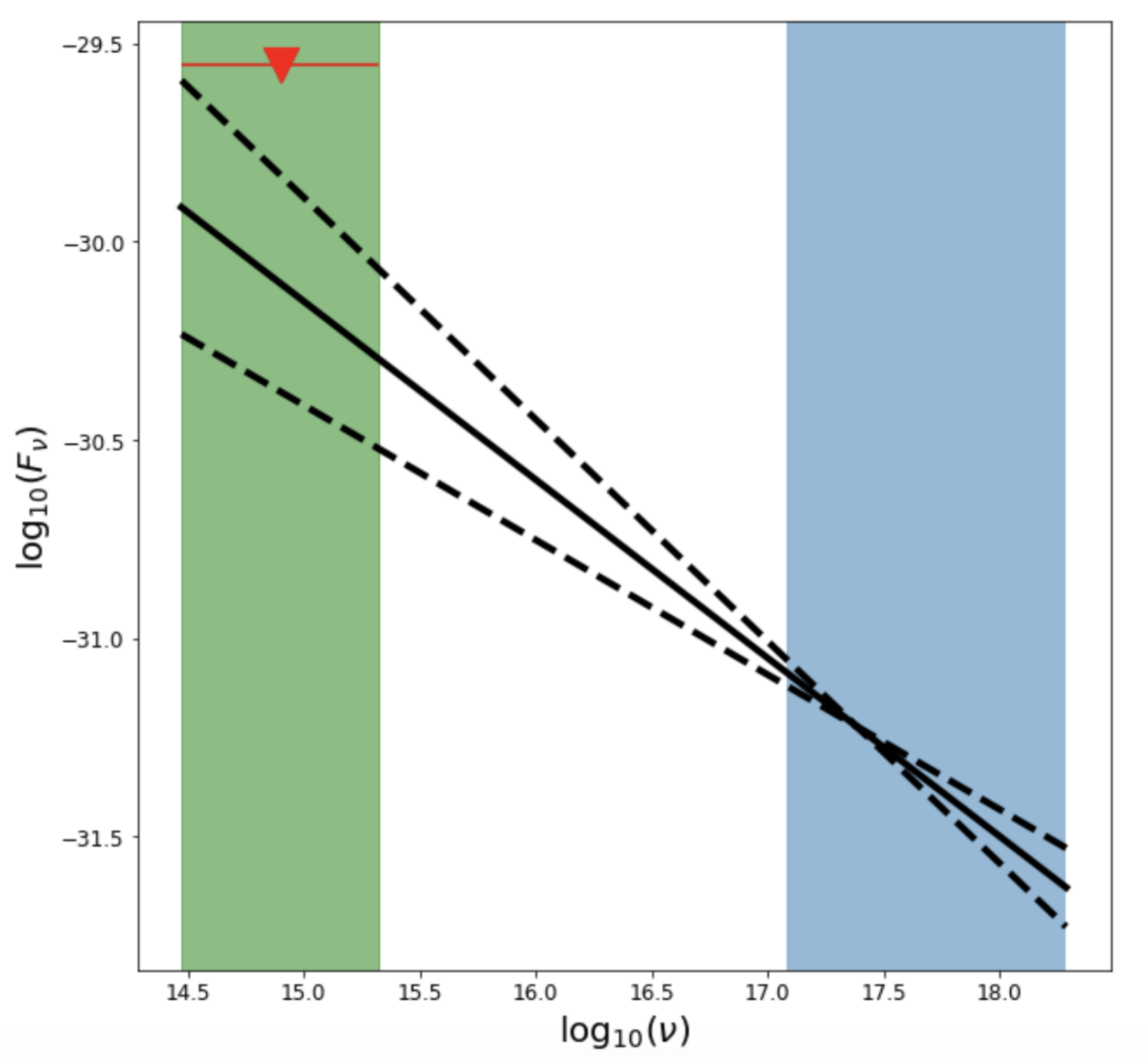}
\caption{Unabsorbed X-ray spectrum (with $\alpha_\nu=0.45\pm0.11$; $F_\nu\propto\nu^{-\alpha_\nu}$)  extrapolated to optical frequencies, along with the {\sl HST} $3\sigma$ upper limit (red triangle). The green and blue bands show the optical and X-ray coverage of {\sl HST} STIS and {\sl CXO}, respectively.
\label{hstspec}
}
\end{figure}

  Detecting the clump at lower frequencies could constrain the broad-band SED shape and
provide a more precise measurement of the spectral slope
(if the SED has a power-law shape). Unfortunately, we did not detect the clump in the {\sl HST} images, with the upper limit being above the continuation of the PL spectrum with the slope inferred from the X-ray spectra (see Figure \ref{hstspec}). 

\subsection{Clump's dynamics: Accelerated Motion}

The data from our earlier campaign corresponding to the preceding binary cycle provided only marginal evidence of the clump being accelerated, assuming that it was launched with a small (compared to the observed) velocity near periastron passage.  P+15  speculated that the PW ram pressure could be responsible for the rapid acceleration. Subsequent numerical simulations  confirmed that, indeed,  the clump can be accelerated to large  velocities by the PW in the $\eta<1$ scenario (see Figure 2 in \citealt{2016MNRAS.456L..64B}).  However, the new data reported here  provide more direct evidence of accelerated motion,
 with the acceleration continuing to be present  on timescales of  hundreds of days after the clump's launch near the periastron passage (see Figure \ref{pic4}). 
   
    Following P+15,  the acceleration can be estimated as $a_{\perp}\sim {\cal F}_{\rm pw}m_{cl}^{-1}$, 
 where ${\cal F}_{\rm pw}=p_{\rm pw}A=(\dot{E}/c)\xi_A\xi_\Omega=2.8\times10^{25}\xi_A\xi_\Omega$ dyn is the PW ram force, $\xi_{A}<1$ is a filling factor (which takes into account that the clump can consist of separate fragments), $\xi_\Omega$ is the fraction of PW that interacts with the clump (see above), and $m_{cl}$ is the mass of the clump.  Using the estimated  $a_{\perp}\approx 50$ cm s$^{-2}$,  we obtain an upper limit on the clump's mass, $m_{cl}\lesssim6\times10^{23}\xi_A\xi_\Omega$ g, and  kinetic energy, $E_{\rm kin}\lesssim6\times10^{42}(m_{cl}/6\times10^{23}$ g)$(v/0.15c$)$^{2}$ erg, at the time of the last observation. This upper limit corresponds to the pulsar's rotational energy losses over the period of $\simeq 80$ days, which is comparable to the time that pulsar spends interacting with the disk (see e.g., Figure 3 in  \citealt{2006MNRAS.367.1201C}).

The above estimate makes the assumption that the accelerating force is due to the pressure of the PW. However, the pressure of the radiation produced by the luminous companion star could represent  an additional  force. To understand the relative contribution of the radiation force,  let us first consider   non-relativistic electrons that can be associated with the ejected fragment of the decretion disk (assuming it is ionized). In this case, the X-ray emission  can  still be synchrotron emission  attributed to a small number of ultra-relativistic electrons accelerated in the shock at the interface between the PW and the clump.

 For Thomson scattering,  the average optical thickness of the clump can be estimated as $\tau\sim n_e\sigma_T r_{\rm cl}\approx1\times 10^{-9}(m_{\rm cl}/10^{26}~{\rm g})(r_{\rm cl}/10^{17}~{\rm cm})^{-2}$,  where 
 $n_e=0.014 (m_{\rm cl}/10^{26}~{\rm g})r_{17}^{-3}$ cm$^{-3}$   is the number density of non-relativistic electrons (for scaling purposes we have conservatively assumed a very large density, corresponding to a proton-electron clump as massive as $10^{26}$ g -- the upper bound on the decretion disk mass; see  above for a more realistic upper limit on the clump's mass). 
If we conservatively assume that in a single photon-electron interaction the entire momentum of the photon can be transferred to the electron (in the electron's rest frame), then the maximum contribution of the radiation force to the acceleration of the clump is ${\cal F}_{\rm rad}\sim \tau \xi^{*}_{\Omega} L_*/c$, where $\xi^{*}_{\Omega}=(r_{\rm cl}/2r)^{2}\approx0.04$ is the fraction of stellar photons intercepted by the clump\footnote{Although in the case of $\eta>1$ we expect $\xi^{*}_{\Omega}$ to be similar to $\xi_{\Omega}$ (introduced above for the PW),  in the opposite, $\eta<1$, case $\xi_{\Omega}\sim1$ while  $\xi^{*}_{\Omega}$  remains the same as in the $\eta>1$ case.}. Therefore, although $L_*/\dot{E}\approx 300$, the ratio of the radiation 
force to the PW force is only  
${\cal F}_{\rm rad}/{\cal F}_{\rm pw}\sim 1\times 10^{-8} \xi_{A}^{-1}\xi_{\Omega}^{-1}(\xi^{*}_{\Omega}/0.04)(m_{\rm cl}/10^{26}~{\rm g})(r_{\rm cl}/10^{17}~{\rm cm})^{-2}$, demonstrating that even in the most optimistic case,
 the radiation pressure exerted on the clump's non-relativistic electrons has a negligible effect on its acceleration.

 Let us now consider the effects of the radiation pressure on the 
relativistic electrons whose  $\gamma$ should  reach $10^{7}$--$10^{8}$ 
(well in excess of $\gamma_{\rm KN}\sim 10^4$)
to produce synchrotron emission in X-rays (see above). 
According to \cite{1981ApJ...243L.147O}, in the Thomson regime ($\gamma\ll\gamma_{\rm 
KN}$) the radiation force exerted on isotropically distributed relativistic electrons with a Lorentz 
factor $\gamma\gg 1$ is a factor of $(2/3)\gamma^2$  larger than in the non-relativistic case, which could lead to rapid acceleration of a clump containing relativistic particles (the ``Compton rocket effect'').
However, in the KN regime, the amplification factor grows with $\gamma$ much more slowly, $\sim 
\gamma_{\rm KN}^2 [\ln(2\gamma/\gamma_{\rm KN}) -5/6]$ for $\ln(\gamma/\gamma_{\rm KN}) \gg 1$ \citep{1974ApJ...188..121B}. To estimate the force exerted on a clump with
 relativistic
 electrons, the force at a given $\gamma$ should be integrated with a power-law SED,
$dN_e/d\gamma = K_e\gamma^{-p}d\gamma$  ($\gamma_{\rm m}<\gamma<\gamma_{\rm M}$). Taking into account that $\gamma_M\gg \gamma_{\rm KN}$ in our case, there is always a contribution from IC scattering in the KN regime, $\propto \gamma_{\rm KN}^2\,[{\rm max}(\gamma_{\rm KN},\gamma_m)]^{1-p}$ (for
$p>1$ and $\gamma_M \gg \gamma_m$), while the Thomson 
regime term, $\propto \gamma_{\rm KN}^{3-p}$ (for $p<3$), only contributes
 at $\gamma_m\ll \gamma_{\rm KN}$. 
Thus, at a given $\gamma_{\rm KN}$ and $1<p<3$, 
the radiative force does not depend on
$\gamma_m$ at $\gamma_m\ll \gamma_{\rm KN}$ and decreases $\propto (\gamma_{\rm KN}/\gamma_m)^{p-1}$ at $\gamma_m\gg \gamma_{\rm KN}$. 

Assuming $\gamma_m\ll \gamma_{\rm KN}$ and expressing the electron
SED normalization in terms of X-ray synchrotron luminosity,
$K_e = L_X \left[(4/3)\sigma_T c u_B C_{3-p}(\gamma_1,\gamma_2)\right]^{-1}$ (see Section \ref{rad_mech}),  we obtain an order-of-magnitude estimate for the radiative force
\begin{equation}
{\cal F}_{\rm rad} \sim q \frac{L_X}{c} \frac{\gamma_{\rm KN}^{3-p}}{C_{3-p}(\gamma_1,\gamma_2)}\,,
\end{equation}
where $q=u_{\rm rad}/u_B$. The ratio of this force to the PW ram force is ${\cal F}_{\rm rad}/{\cal F}_{\rm pw} \sim (\xi_X/\xi_A) q \gamma_{\rm KN}^{3-p}/C_{3-p}(\gamma_1,\gamma_2)$, 
where $\xi_X = L_X/(\xi_\Omega\dot{E})$ is the X-ray efficiency (see Section \ref{rad_mech}).
For $p=1.85$, we obtain
${\cal F}_{\rm rad}/{\cal F}_{\rm pw} \sim {\rm a\,\,few} \times 10^{-6}
(\xi_X/10^{-3.5}) \xi_A^{-1} r_{17}^{-2} (\gamma_{\rm KN}/10^{4})^{1.15} (B/100\,\mu{\rm G})^{-1.425}$. This ratio becomes even smaller (by a factor of $(\gamma_m/\gamma_{\rm KN})^{0.85}$) if $\gamma_m\gg \gamma_{\rm KN}$.
 Thus, although the mean radiative force per electron is larger
than in the nonrelativistic case, the number of relativistic electrons required to provide the observed synchrotron luminosity is relativly small, so the
contribution of the radiative force to the clump acceleration is negligible.

\subsection{Comparison to the previous binary cycle\footnote{See also \citet{2019arXiv190300781P}. }}

The differences in the clump properties between the two binary cycles could be related  to the difference in the GeV flares, if the flares are associated with the disk fragmentation and  the destroyed disk fragments can be considered as clump ``seeds''. The different structure of the 2010 and 2014 GeV light curves likely reflects different masses and initial speeds of the disk fragments, which would ultimately result in different clumps. For instance, the later appearance
of the resolved clump and longer acceleration time (but about the same  final velocity) in the 2014\ndash2017 binary cycle could be linked to the flatter  GeV flare light curve caused by a longer and more gradual process of disk destruction and a larger mass of the fragment \citep{2015ApJ...811...68C}.
 Since the 2017 GeV flare light curve \citep{2018ApJ...863...27J} was very different from the 2010 and 2014 light curves (larger  fluence, longer delay, higher peak), we can expect an even larger difference in the clump properties in the 2017\ndash2021 binary cycle. Future {\sl CXO} observations will test this hypothesis.

 The behavior of the extended X-ray emission  around B1259 
in the 2014\ndash2017 cycle  demonstrated several other potentially important differences with respect to what was seen in the 2010\ndash2014 cycle (P+15;
\citealt{2019arXiv190300781P}). In the image from ObsID 19270 (see Figure \ref{pic2}),  the clump's morphology evolved into  an  interesting linear extended feature, nicknamed the ``whiskers''.  This change in the clump's shape (the whiskers are not visible in the next observation $\approx100$ days later,  which is roughly the light travel time along the structure)  was not seen in the previous cycle, but it may have been missed since it is not a persistent feature.  The ``whiskers'' show that the width of the channel occupied by the PW which pushed the clump (our preferred scenario) must be rather large at these distances. In the following observation (108 days after the one where the whiskers were detected) the clump 
appeared to be more compact but underwent an episode of  re-brightening, a behavior that was also not previously seen.

A qualitative picture explaining the whiskers and brightening could be as follows. As the pulsar passes through periastron, some amount of stellar wind becomes entrained into the PW.\footnote{This could happen either because the stellar wind is collimated in the apastron direction (if $\eta>1$) or (if $\eta<1$) because the channel carved out by the PW allows some stellar wind particles to leak into the channel.} Then, the pulsar interacts with the companion's decretion disk, fragmenting it and launching the clump. In the $\eta<1$ scenario, this clump will then sweep up the stellar wind particles entrained in the channel as it is being accelerated by the PW ram pressure.  The ``whiskers'' could be outflows from the clump, which is flattened by the combination of the PW ram pressure and the drag force pressure from the residual stellar matter in the channel.  At some location along the channel,  the medium will transition from being dominated by the PW into a mixture of the pulsar and stellar winds (see e.g., Figure 1 in \citealt{2016MNRAS.456L..64B}),
with a lower sound speed. 
Therefore, as the clump travels farther away from the pulsar, at some point its speed may become supersonic. This transition could result in a shock, which could give rise to the observed brightening (e.g., if the magnetic field is compressed), and the bow-shock shaped morphology of the clump. The whiskers and the brightening episode are more difficult to explain in the $\eta>1$ scenario. Alternatively, as briefly mentioned in Section \ref{rad_mech}, the rather hard SED slope ($p=1.45$) and a lack of synchrotron cooling could suggest that the emitting particles are being re-accelerated via magnetic reconnection, which could in principle also contribute to the acceleration of the bulk flow (see e.g., \citealt{2001ApJ...547..437L,2017MNRAS.468.3202B}). The inherently variable nature of magnetic reconnection may also explain the sudden brightening of the clump seen in the ObsID 20054 image. However, more observations with a higher cadence are necessary to probe these scenarios. 

During the previous cycle, the the clump showed a steadily decaying flux as it traveled further away from the binary  (see Figure \ref{pic10}). The decreasing flux of the clump may be  attributed to the decreasing magnetic field as a function of distance from the pulsar in the synchrotron emission scenario. In this case the flux should be $\propto B^{(p+1)/2}$. Alternatively, if the clump is confined to a channel which is diverging rapidly at the distances where we observe the clump, adiabatic cooling may play a noticeable role.

Finally, there is solid evidence for a separate extended feature (second clump) $\simeq2\farcs2$ away from the binary core in the image from  ObsID 20054. However,  it was not detected in the next observation. This suggests that it faded on timescales shorter than 120 days. Unfortunately, the second feature is too faint and too close to  the bright binary to extract a reliable spectrum. However, we have revisited our older observations near this orbital phase (i.e., ObsIDs 16553+16583 occurring 1151 days after periastron passage, see P+15 for additional details) and found evidence of a faint second feature located 3\farcs4 from the binary (see Figure \ref{pic13}). The feature has $\sim30$ net counts in the combined ACIS image.
 The spectrum can be fitted by an absorbed PL  model with $\Gamma=1.6\pm0.4$ and unabsorbed flux $f_X=1.1(2)\times10^{-14}$ erg cm$^{-2}$ s$^{-1}$ for fixed $N_{\rm H}=3\times10^{21}$ cm$^{-2}$. The second clump has a similar photon index and flux as the first clump, but these values are not well constrained due to the small number of counts. Interestingly, this second (smaller) clump is launched in the same direction as the first (larger) clump in both orbital cycles. 
In order to understand the nature of these differences between the two different binary cycles, continued monitoring of this system with {\sl CXO} is needed.

\section{Conclusions}
\label{summ}

The new {\sl CXO} observing campaign of B1259 during the 2014-2017 orbital cycle has allowed us to confirm the recurrent (with every binary cycle)  nature of the remarkable high-velocity features (clumps)
 ejected from the binary. Although there were a lot of similarities with the previous binary cycle, there were also some important differences. The feature appears to detach from the binary later (i.e., more time has passed since periastron passage) than in the previous binary cycle. If the clump is ejected near periastron passage, it must be  accelerated up to the observed projected velocity,
$ v_{\perp}\approx 0.15c$, with a projected acceleration, $a_{\perp}\approx50$ cm s$^{-2}$. The evidence for acceleration is present at large distances (i.e., a few arcseconds) from the binary. The   acceleration measurement provides an upper limit on the mass ($m_{\rm cl}\lesssim 6\times10^{23}$ g)  and kinetic energy  ($E_{\rm kin}\lesssim6\times10^{42}$ erg) of the clump.
 We find that the radiation pressure force is not sufficient to continue accelerating the clump to the observed speed at the observed separations from the star. Therefore, the accelerating force  is likely provided by PW.
   The observed X-ray emission is likely to be synchrotron radiation from the interface between the clump material and PW.
  Further, we show that IC upscattering is unlikely to be responsible for the X-ray emission.

The clump showed puzzling changes in morphology (developing  short-lived ``whiskers'' in one of the images). Unlike the  clump observed in the 2010-2014 orbital cycle, this clump shows an episode of brightening. In the $\eta<1$ scenario, these whiskers can be explained if the clump sweeps up some stellar matter that somehow leaks into the channel, causing a flattening in the clump due to the the PW ram pressure pushing the clump, and the drag from the stellar wind in front of the clump. The brightening may be explained if the clump is supersonic as it crosses the region between the PW channel, and the region filled with a mixture of pulsar and stellar wind. We also find evidence of a second (more transient) clump launched in both cycles after the first more prominent clump. While we do not find direct observational support for either the $\eta>1$ or $\eta<1$ scenario,  the observed variability in morphology and brightness could be easier explained in the $\eta<1$ scenario. Future {\sl CXO} observations will allow us to probe the possible connections between the GeV flares and phenomenology of the X-ray emitting clump.

\acknowledgements
Support  for  this  work  was provided by the National Aeronautics and Space Administration
through Chandra Awards  GO5-16065, GO7-18056 and DD7-18088,
  issued  by  the Chandra
X-ray Observatory Center, which is operated by the
Smithsonian Astrophysical Observatory for and on behalf of the
National Aeronautics and Space Administration under contract
NAS8-03060. Support for {\sl HST} program 14932 was  provided by NASA through a
grant  from the Space Telescope Science Institute, which
is  operated  by  the  Association  of  Universities  for  Research  in  Astronomy,  Incorporated,  under  NASA  contract NAS5-26555.  JH's work was also supported by the George Washington University  through the funding made available to the Astronomy, Physics, and Statistics Institute of Sciences (APSIS). JH and OK would like to thank Wenbin Lu for useful discussions. 
We are grateful to the CXC Director Belinda Wilkes
and STScI Director Kenneth Sembach for allocation the {\sl CXO} and {\sl HST} DDT observations.

\end{document}